\documentclass{iopart}
\usepackage{iopams}
\usepackage{setstack}
\usepackage{graphicx}
\usepackage{bm}
\usepackage[usenames]{color}

\newcommand{\be}{\begin{equation}}
\newcommand{\ee}{\end{equation}}
\newcommand{\beqn}{\begin{eqnarray}}
\newcommand{\eeqn}{\end{eqnarray}}

\begin{document}

\title[Critical quench dynamics of random quantum spin chains]{Critical quench dynamics of random quantum spin chains:
Ultra-slow relaxation from initial order and delayed ordering from initial disorder}

\author{Gerg\H o Ro\'osz$^{1,2}$, Yu-Cheng Lin$^{3}$, Ferenc Igl\'oi$^{1,2}$}
\address{$^1$ Wigner Research Centre, Institute for Solid State Physics and Optics,
H-1525 Budapest, P.O.Box 49, Hungary}
\address{$^2$ Institute of Theoretical Physics,
Szeged University, H-6720 Szeged, Hungary}
\address{$^3$Graduate Institute of Applied Physics,
National Chengchi University, Taipei, Taiwan}
\ead{roosz.gergo@wigner.mta.hu}
\ead{yc.lin@nccu.edu.tw}
\ead{igloi.ferenc@wigner.mta.hu}

\begin{abstract}
By means of free fermionic techniques combined with multiple precision arithmetic we
study the time evolution of the average magnetization, $\overline{m}(t)$, of
the random transverse-field Ising chain after global quenches. 
We observe different relaxation behaviors for quenches starting
from different initial states to the critical point. 
Starting from a fully ordered initial state,
the relaxation is logarithmically slow described by  $\overline{m}(t) \sim \ln^a t$,
and in a finite sample of length $L$ the average magnetization saturates at a size-dependent plateau 
$\overline{m}_p(L) \sim L^{-b}$; here the two exponents satisfy the relation
$b/a=\psi=1/2$. Starting from a fully disordered initial state,
the magnetization stays at zero for a period of time until $t=t_d$ with $\ln t_d \sim L^{\psi}$ 
and then starts to increase until it saturates to an asymptotic value $\overline{m}_p(L) \sim L^{-b'}$, 
with $b'\approx 1.5$. For both quenching protocols, 
finite-size scaling is satisfied in terms of the scaled variable $\ln t/L^{\psi}$.        
Furthermore, the distribution of long-time limiting values of the magnetization
shows that the typical and the average values scale differently
and the average is governed by rare events. The non-equilibrium
dynamical behavior of the magnetization is explained through semi-classical theory.
\end{abstract}

\maketitle
\section{Introduction}
\label{sec:intr}

Following the experimental progress in non-equilibrium dynamics of
ultracold-atomic gases in optical
lattices~\cite{Greiner_02,Paredes_04,Kinoshita_04,Sadler_06,Lamacraf_06,Kinoshita_06,Hofferberth_07,bloch,Trotzky_12,Cheneau_12,Gring_11},
there are tremendous theoretical efforts aimed at understanding the
time-evolution of certain observables in closed quantum systems after a sudden
or smooth change of Hamiltonian parameters. In a quench process both the functional
form of the relaxation and the properties of the long-time, presumably
stationary state are of
interest~\cite{Polkovnikov_11,barouch_mccoy,igloi_rieger,sengupta,Rigol_07,Calabrese_06,Calabrese_07,Cazalilla_06,Manmana_07,
Cramer_08,Barthel_08,Kollar_08,Sotiriadis_09,Roux_09,Sotiriadis_11,Kollath_07,Banuls_11,Gogolin_11,Rigol_11,Caneva_11,
Rigol_12,Santos_11,Grisins_11,Canovi_11,Calabrese_05,Fagotti_08,Silva_08,Rossini_09,Campos_Venuti_10,Igloi_11,Rieger_11,Foini_11,
Calabrese_11,Schuricht_12,Calabrese_12,blass,Essler_12,evangelisti_13,fagotti_13,pozsgai_13a,fagotti_essler_13,collura_13,
bucciantini_14,fagotti_14,cardy_14,wouters,pozsgay,goldstein,pozsgay1,pozsgay2,larson,hamazaki,blass_rieger,essler_mussardo_manfil,ilievski1,ilievski2,doyon,vidmar}.
In homogeneous systems the order parameter has an exponential relaxation, while
the entanglement entropy between a subsystem and its environment grows linearly
in time after a quench. These phenomena have been explained in the
frame of a semi-classical theory. According to the theory, during the quench process
quasi-particles are created uniformly in the sample and  move
ballistically~\cite{Calabrese_05, Calabrese_07,Igloi_11,Rieger_11}. Concerning
the long-time limit of the relaxation process integrable and non-integrable
systems show different behaviors. Non-integrable systems are expected to
thermalize~\cite{Rigol_07,Calabrese_06,Calabrese_07,Cazalilla_06,Manmana_07,
Cramer_08,Barthel_08,Kollar_08,Sotiriadis_09,Roux_09,Sotiriadis_11}, 
but there are some counterexamples~\cite{larson,hamazaki,blass_rieger}. On the other hand,
integrable systems in the stationary state are generally described by a
so-called generalized Gibbs ensemble, which includes all the conserved
quantities of the system. Recently it has been observed that in certain models
both local and quasi-local conserved quantities
have to be
taken into account to construct an appropriate
generalized Gibbs ensemble~\cite{wouters,pozsgay,goldstein,pozsgay1,pozsgay2,essler_mussardo_manfil,ilievski1,ilievski2,doyon,vidmar}.

In a system with spatial inhomogeneities, such as
defects~\cite{peschel03,PeschelZhao,Levine08,ISzL09,Eisler_Peschel10,CMV11,peschel12},
quasi-periodic or aperiodic interactions~\cite{IJZ07,Igloi_13,Igloi_14}, the
non-equilibrium relaxation process becomes qualitatively different from the homogeneous case. In the
semi-classical picture the quasi-particles are also inhomogeneously created and
they move in a complex diffusive way. With randomness the eigenstates of
the Hamiltonian can be localized, which leads to Anderson
localization~\cite{anderson} in non-interacting systems or many-body
localization~\cite{mbl} if the particles are interacting; in these cases 
the quasi-particles generally move only finite distances away from
their place of creation  or follow some ultra-slow diffusive motion. As a result
the non-equilibrium long-time stationary state of disordered systems retains 
memory of the initial state and therefore thermalization does not take
place~\cite{huse}.

Concerning the functional form of the relaxation process after a quench in
random quantum systems, there have been detailed studies about the time-dependence of
the entanglement entropy~\cite{dyn06,Igloi_12,Levine_12,Pollman_12,sirker}. If
the system consists of non-interacting fermions - such as the critical XX-spin
chain with bond disorder or the critical random transverse-field Ising chain - the
dynamical entanglement entropy grows ultraslowly in time as 
\be
 {\cal S}(t) \sim a \ln \ln t\,,
 \label{eq:S_lnln}
\ee
and saturates in a finite system at a value 
\be
   {\cal S}(\ell) \sim b \ln \ell\,,
 \label{eq:S_longt}  
\ee 
where $\ell$ denotes the size of a block in a bipartite system and can be chosen 
to be proportional to the size of the system $L$~\cite{Igloi_12,sirker}. These scaling forms
can be explained by a strong disorder renormalization-group (SDRG)
approach~\cite{im}. Recently, the SDRG method, which was designed as a ground state
approach, has been generalized to take into account excited
states~\cite{pekker,Vosk_13,Vosk_14}; this generalized RG method is often
abbreviated as RSRG-X~\cite{pekker}.  By this generalized SDRG method the ratio
of the prefactors in (\ref{eq:S_lnln}) and (\ref{eq:S_longt}) is
predicted as $b/a=\psi_{\mathrm{ne}}$, where $\psi_{\mathrm{ne}}=1/2$ is a critical exponent
in the non-equilibrium process and describes the relation between
time-scale and length-scale as %
\be
\ln t \sim L^{\psi_{\mathrm{ne}}}\;.
\label{psi_ne}
\ee
For interacting fermion models due to many-body localization the
time-dependence of the dynamical entropy is ${\cal S}(t) \sim  \ln^{\omega} t$
with $\omega \ge 1$, while the saturation value follows the volume law, 
${\cal S}(\ell) \sim \ell$~\cite{Pollman_12}.

In the present paper we study the relaxation of the order parameter of the
random transverse-field Ising chain after a quench to the critical final state
and into a ferromagnetic state.  We consider relaxation processes from an
initial ferromagnetic state and from a fully paramagnetic state by a sudden
change of the strength of the transverse field.  To circumvent numerical
instability as observed in previous calculations for large systems 
using eigenvalue solver routines~\cite{Igloi_12,sirker}, we use
multiple precision arithmetic to study the time-evolution through direct matrix
multiplications.

The rest of the paper is organized as follows. The model and the method for
calculating the local magnetization are described in section~\ref{sec:model} . The
numerical results for relaxation of the magnetization following 
different quench protocols are presented and discussed in section~\ref{sec:relax}.
A summary is given in section~\ref{sec:disc}. Details of the time evolution of Majorana fermion
operators are given in the appendix.

\section{The model and the method}
\label{sec:model}

The model we consider is the random transverse-field Ising chain of length $L$ defined by the Hamiltonian:
\be
{\cal H}=-\frac{1}{2}\sum_{i=1}^{L-1} {J}_i \sigma_i^x {\sigma}_{i+1}^x 
  -\frac{1}{2}\sum_{i=1}^{L} {h}_i {\sigma}_i^z\;,
\label{hamilton}
\ee
in terms of the Pauli matrices ${\sigma}_i^{x,z}$ at site $i$. In this paper we will
consider open chains with free boundary conditions.
The couplings, ${J}_i$, and the transverse fields, ${h}_i$, are position
dependent random numbers taken from the uniform distributions
in the intervals $[0,1]$ and $[0,1]h$, respectively.
The strength of the random transverse field, $h$, is time-dependent;
for $t<0$ its value is $h_0$ and for $t > 0$ it changes suddenly to $h\,(\neq h_0)$.
The initial and the final Hamiltonians are denoted by ${\cal H}_{0}$
and ${\cal H}$, respectively. The system for $t<0$ is prepared
in the ground state $|\Psi_0^{(0)}\rangle$ of the initial Hamiltonian; 
after the quench for $t > 0$ the system evolves according to 
the final Hamiltonian ${\cal H}$ and the state of the system 
is described by $|\Psi_0(t)\rangle=\exp(-\imath{\cal H}t)|\Psi_0^{(0)}\rangle$,
which is generally not an eigenstate
of ${\cal H}$. Here and throughout the paper we denote the imaginary
unit $\sqrt{-1}$ by $\imath$ to avoid confusion with the integer index $i$ and
set $\hbar=1$. To calculate the time-dependent expectation value $A(t)$ of an observable
$\hat{A}$, we work in the Heisenberg picture using $\hat{A}_H(t)=\exp(\imath{\cal H}t)\hat{A}\exp(-\imath{\cal H}t)$
and evaluate $A(t)=\langle \Psi_0^{(0)}|\hat{A}_H(t)|\Psi_0^{(0)}\rangle$.
Time-dependent matrix elements and correlation functions are calculated 
in a similar way.

The standard way to deal with the Hamiltonian of the transverse field Ising chain 
is the mapping to spinless free fermions~\cite{lieb61,pfeuty}.
The spin operators
${\sigma}_i^{x,y,z}$ are expressed in terms of fermion creation (annihilation) operators
${c}_i^\dagger$ (${c}_i$) by using the Jordan-Wigner
transformation~\cite{JW}:  ${c}^\dagger_i={a}_i^+\exp\left[\pi \imath \sum_{j}^{i-1}{a}_j^+{a}_j^-\right]$
and ${c}_i=\exp\left[\pi \imath
\sum_{j}^{i-1}{a}_j^+{a}_j^-\right]{a}_i^-$, where ${a}_j^{\pm}=({\sigma}_j^x \pm \imath{\sigma}_j^y)/2$.
The Ising Hamiltonian in~(\ref{hamilton}) can then be written in a quadratic form in fermion operators:
\be
  {\cal H}=-\sum_{i=1}^{L}h_i \left( {c}^\dagger_i {c}_i-\frac{1}{2} \right)\nonumber -
\frac{1}{2}\sum_{i=1}^{L-1} J_i({c}^\dagger_i-{c}_i)({c}^\dagger_{i+1}+{c}_{i+1})\;.
\label{ferm_I}
\ee
In this paper we are interested in the relaxation of the local order parameter
(magnetization), $m_l(t)$, at a position $l$.  Following the method by
Yang~\cite{Yang_52}, for a given sample of large length $L$ we define $m_l(t)$
as the off-diagonal matrix element of the longitudinal magnetization operator:
$m_l(t)=\langle \Psi_0^{(0)}|{\sigma}_l^x(t)|\Psi_1^{(0)}\rangle$, where
$|\Psi_1^{(0)}\rangle$ is the first excited state of ${\cal H}_{0}$. In the
free-fermion representation we introduce two Majorana fermion operators,
$\check{{a}}_{2i-1}$ and $\check{{a}}_{2i}$ at site $i$ with the definition:
\beqn
\check{{a}}_{2i-1}={c}^{\dag}_i+{c}_i\nonumber \\
\check{{a}}_{2i}=-\imath({c}^{\dag}_i-{c}_i)\,,
\label{maj_def}
\eeqn
which obey the commutation relations:
\be
\check{{a}}_{i}^{+}=\check{{a}}_{i},\quad \{ \check{{a}}_{i},\check{{a}}_{j} \}=2\delta_{i,j}\;.
\label{comm_maj}
\ee
The spin operators are then expressed in terms of the Majorana operators as:
\begin{eqnarray}
{\sigma}_l^x&=&\;\imath^{l-1} \prod_{j=1}^{2l-1} \check{{a}}_{j} \;.
\label{sigma_xz}
\end{eqnarray}
The calculation of the magnetization, $m_l(t)$, involves the evaluation of
the determinant of a $2l \times 2l$ antisymmetric matrix ${\bf C}$ with the
elements being the correlation functions $C_{ij}=\langle \mathit{\Psi}^{(0)}_0
|\check{{a}}_{i}(t)\check{{a}}_{j}(t) | \mathit{\Psi}^{(0)}_0 \rangle$, for
$i<j<2l$, $C_{i,2l}=\langle \mathit{\Psi}^{(0)}_0 |\check{{a}}_{i}(t) |
\mathit{\Psi}^{(0)}_1 \rangle$, and $C_{ji}=-C_{ij}$. For details see the
appendix of Ref.~\cite{Igloi_13}. 

For a random system any physical quantity requires an averaging over disorder realizations.
We denote the disorder average by an overbar; accordingly, the average local magnetization
is expressed as 
\be
  \overline{m}_l(t)=\overline{\langle \Psi_0^{(0)}|{\sigma}_l^x(t)|\Psi_1^{(0)}\rangle}\,.
\ee

\subsection{In equilibrium}

In equilibrium (i.e. for $t < 0$), the critical behavior of the local magnetization $m_l(0)$ has been
analytically studied by the SDRG method~\cite{fisher}, and the random quantum
critical point (at $h_0=h_c=1$) is found to be controlled by an infinite-randomness fixed
point~\cite{danielreview,im}.  In the thermodynamic limit ($L\to \infty$) one
needs to discriminate between bulk ($l/L={\cal O}(1)$) and surface ($l={\cal O}(1)$) points, 
and the corresponding average magnetization is denoted by
$\overline{m}$ and $\overline{m^s}$, respectively. In the ordered phase,
$h_0<h_c=1$, we have $\overline{m}>0$, which vanishes at the critical point as
$\overline{m} \sim (h_c-h_0)^{\beta}$ with $\beta=(3-\sqrt{5})/2$. The surface
magnetization follows similar behavior, however with a surface exponent
$\beta_s=1$. At the critical point of a finite system the scaling of the
average magnetization is governed by rare samples (or rare regions), which have
the magnetization of order of ${\cal O}(1)$, whereas in typical samples the magnetization
behaves as $m_{\mathrm{typ}} \sim \exp(-A\sqrt{L})$. 
The fraction of rare events scales as $P_{\mathrm{rare}}\sim L^{-x}$, 
and the same holds for the average magnetization: $\overline{m} \sim L^{-x}$; 
here the exponent $x=\beta/\nu$ is related to the critical
exponent $\nu=2$ of the average correlation function.  
Similarly, the scaling behavior
of the surface magnetization involves the exponent $x_s=\beta_s/\nu$.
Concerning equilibrium dynamical scaling at the infinite-randomness fixed
point, it is extremely space-time anisotropic so that the typical length, $\xi$, and the typical time, $\tau$,
is related as
\be
\ln \tau \sim \xi^{\psi}\;,
\label{psi}
\ee
with an exponent $\psi=1/2$.

\subsection{Out-of-equilibrium}

Some out-of-equilibrium properties of the random transverse-field Ising chain have been predicted
by the RSRG-X approach~\cite{pekker,Vosk_13,Vosk_14}. 
For a quench to the critical state, i.e. $h=h_c$ in the final Hamiltonian,
the relation between the typical length scale and the typical time scale is in the
same form as (\ref{psi}) in the equilibrium case and 
even the corresponding exponents are the same $\psi=\psi_{\mathrm{ne}}=1/2$ (cf. (\ref{psi_ne}) and (\ref{psi})). 
However, available numerical results~\cite{Igloi_12,sirker} for the relaxation of the entanglement
entropy are not in complete agreement with the RSRG-X prediction.

Here we focus on the relaxation of the magnetization in the random chain, which
to our knowledge has not been previously studied. The crucial point of the
method is the calculation of the time evolution of the Majorana operators
introduced in (\ref{maj_def}), which can be given in a closed form, 
provided the Hamiltonian is expressed in a quadratic form in terms of fermion operators,
like the form in (\ref{ferm_I}).
The calculation for  a general quadratic Hamiltonian is described in the appendix. 
Defining $\check{\bf a}(t)$ as a vector with components $\check{a}_{m}(t)$ for
$m=1,2,\dots,2L$, we can express the time evolution of the Majorana operators as
\be
\check{\bf a}(t)={\bf P}(t)\check{\bf a}(0)\;,
\ee
where the $2L \times 2L$ matrix ${\bf P}(t)$ is given by
\be
{\bf P}(t)=\exp({\bf M}t)\;,
\label{P(t)}
\ee 
with an antisymmetric matrix ${\bf M}$ defined in (\ref{M}). 
For our model given in (\ref{ferm_I}) the matrix ${\bf M}$ corresponds to
\begin{equation}
{\bf M}=\left[
 \begin{array}{cccccc}
0   & h_1         &              &        &                  &    \\
-h_1 &    0        & J_1          &        &                  &    \\
    & -J_1         &  0           & h_2    &                  &    \\
    &             &  -h_2         & 0      &    \ddots        &    \\
    &             &              & \ddots & \ddots           & h_L \\
    &             &              &        &  -h_L             & 0
 \end{array}
\right]
\end{equation}
To evaluate the matrix exponential in (\ref{P(t)}), one can use spectral
decomposition of ${\bf M}$ by diagonalizing the large $2L \times 2L$ matrix.
For disordered systems, because of some extremely small eigenvalues standard
eigenvalue solvers would fail to converge for some large-size samples, 
leading to significant numerical errors.  This problem was observed in
our preparatory work. Therefore, we reformulated our numerical procedure to
avoid using any eigenvalue solver routine and instead solve
the time-evolution problem by matrix multiplication using multiple precision arithmetic.
In our numerical procedure we first evaluated the matrix exponential 
at a unit time step, $t_{st}=1$, using the Taylor expansion:
\be
  {\bf P}(t_{st})=\exp({\bf M}t_{st})=\sum_{n=0}^{\infty}\frac{{\bf M}^n}{n!}\,,
  \label{Taylor}
\ee
and calculated the sum of the first hundred terms with multiple precision arithmetic;
the absolute value of the last term of this sum is less than $2^{100}/100!\approx 1.3 \times 10^{-128}$ 
since the eigenvalues of $\bf{M}$ are in the range $[-2,2]$~\cite{deylon}.
The truncation error in (\ref{Taylor}) is therefore sufficiently small even for octuple precision.
For larger time steps we used the identity: ${\bf P}(t_1){\bf P}(t_2)={\bf P}(t_1+t_2)$ 
and iterated it with $t_1=t_2$, starting from $t_1=t_{st}=1$ until a given time step $t_n=2^n$.
We have checked the accumulation of errors by the condition that the matrix
$\bf{P}$ should be orthogonal. 
While calculating the dot products of the arrays of $\bf{P}$, the off-diagonal products have to be zero. 
The sum of the absolute value of the numerically calculated off-diagonal dot products are
found to be less than $10^{-30}$, even after $n=250$ iterations, which
represents the longest time in our calculation.
 
\section{Relaxation of the magnetization}
\label{sec:relax}
In this chapter we present our numerical results for the relaxation of the magnetization
of the random transverse-field Ising chain from two different initial states.
In the first part the initial state is fully ordered with $h_0=0$ for the
parameter of the transverse field; in the second part we consider a
fully disordered initial state with $1/h_0=0$.
Using the method described in section~\ref{sec:model} we have considered
time dependence of the bulk magnetization in finite chains 
of length $L=16,32,64,128$ and $256$ over a time up to $t_{\mathrm{max}}=2^{250}$, i.e.
$\ln(\ln t_{\mathrm{max}})\approx 5.15$. The bulk magnetization $m_l$ in an open chain
was taken in the centre of the chain at $l=L/2$.  
For each system sizes about $10,000$ samples
were considered to obtain the disorder average. 
We have used double-double or quadruple precision arithmetic for our numerical study
and also in some cases checked the accuracy by comparing the results
in the large-time limit with those obtained
with octuple precision.

\subsection{Relaxation from a fully ordered initial state}
\label{sec:ordered_c}
We first consider quenches starting from a fully ordered initial state with $h_0=0$.

\subsubsection{Ferromagnetic final state}


\begin{figure}[t]
\begin{center}
\includegraphics[width=6.3cm]{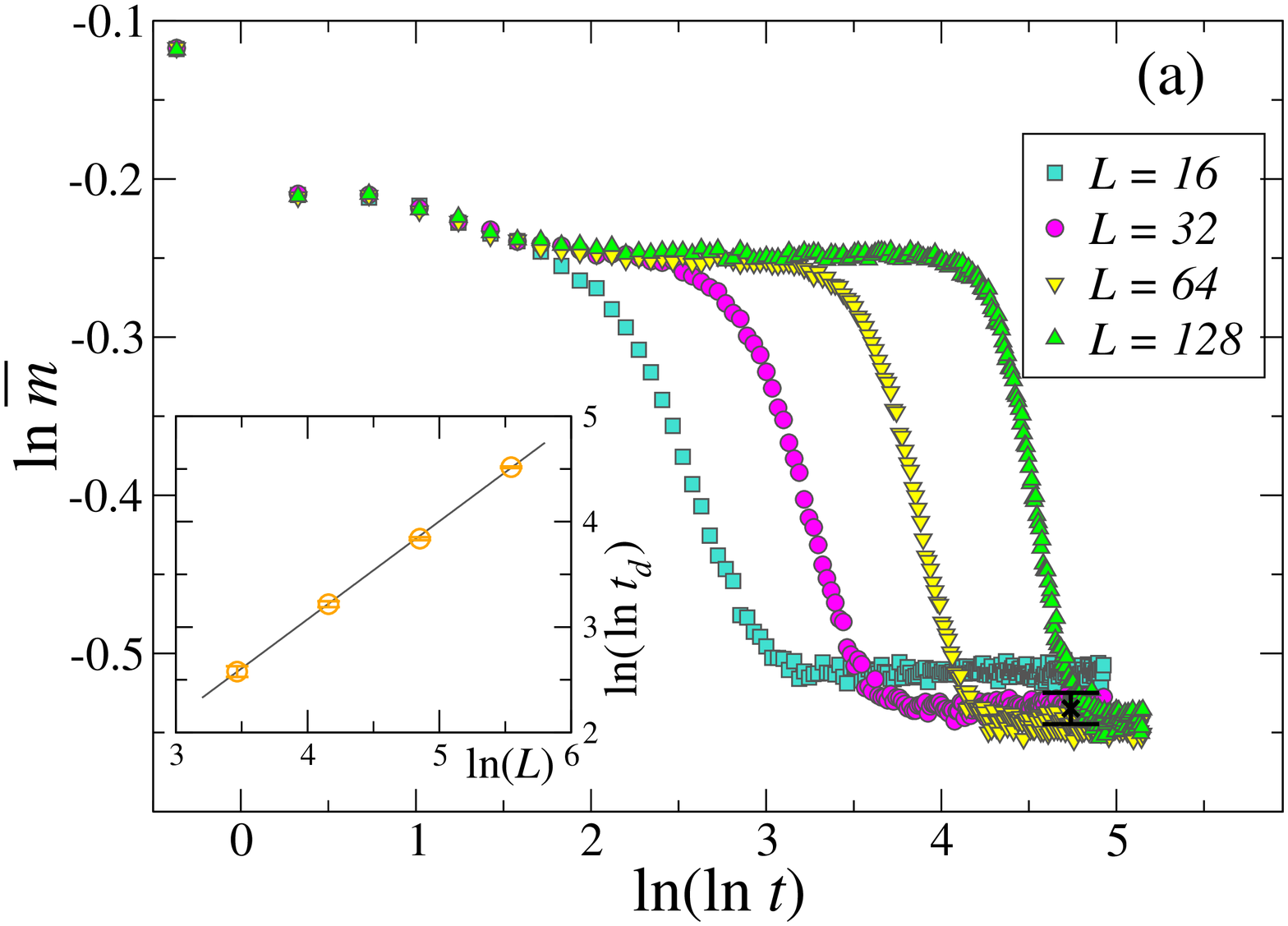}
\includegraphics[width=6.3cm]{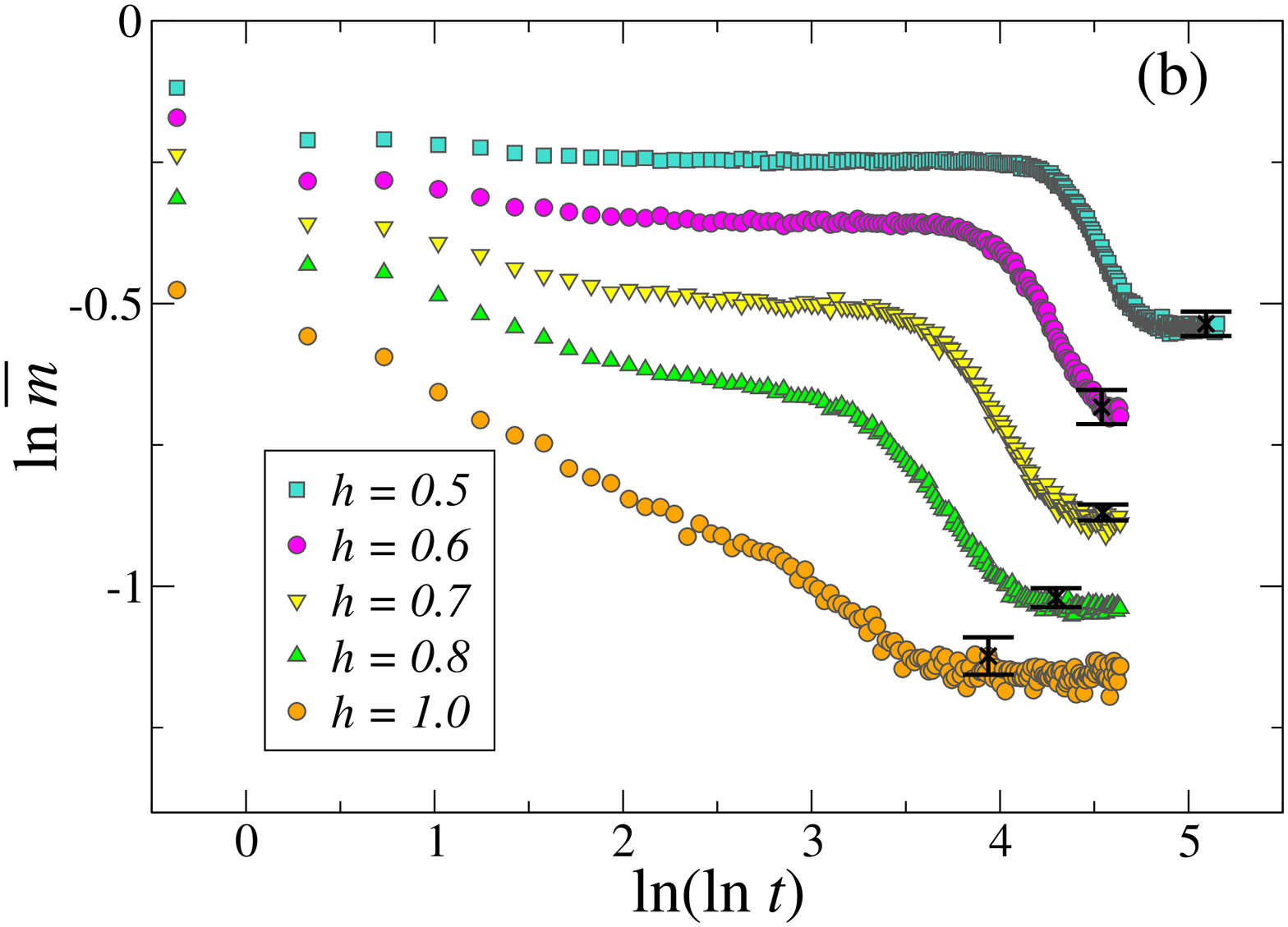}
\end{center}
\caption{Relaxation of the average magnetization after a quench from a fully ordered state to states within the 
ferromagnetic phase. (a) Relaxation for different system sizes with the final parameter of the transverse field $h=0.5$;
the largest error in these data sets is given by the error bar (black).
The inset shows the delay time $t_d$ when the inflection point of the second decay occurs, as a function of the system size $L$; the slope of the straight line is $0.94$.
(b) Relaxation in a chain of length $L=128$ to final states with various values of $h<1$ in the ferromagnetic phase
and also with the critical value $h=h_c=1$; the first plateau seen in the data for quenches to a ferromagnetic phase with small
$1-h$ is smeared out as the final state approaches the critical point with $h=1$. 
The largest errors in each data set are indicated by the error bars.}
\label{fig1}
\end{figure}


Before we study quenches to the critical point, 
we first consider the situation in which the system is quenched to a state within the
ferromagnetic state with $h<1$. In figure~\ref{fig1}(a) we 
present the time dependence of the average magnetization after a quench to $h=0.5$ for different chain lengths
from $L=16$ to $L=128$. As seen in this figure, there is first a decay,
followed by a plateau extending to time $t_d$, after which ($t>t_d$) the magnetization decays sharply before it saturates at a
second plateau; the delay time $t_d$ is $L$-dependent and scales approximately as
$\ln(\ln t_{d}) \approx \ln L+\mathrm{const}$, as shown in the inset of figure~\ref{fig1}(a). 
Similar characteristics is observed for other values of $h<h_c$ (shown in figure~\ref{fig1}(b))  
when the final state is within the ferromagnetic phase.
 
To explain the behavior observed in figure~\ref{fig1} 
we recall that according to a semiclassical theory~\cite{Calabrese_05,Calabrese_07,Igloi_11,Rieger_11}
the local magnetization at a given site decreases in time if 
an odd number of quasi-particles, created by the quench, pass through the site.
In a random chain the motion of the quasi-particles is limited
within a finite localization length $\xi_{\mathrm{loc}}(h)$, which depends on $h$ and grows as the critical point is approached; 
thus the decay of the magnetization stops at some
point of time when the travel distance of the quasi-particles is
reached, which explains the presence of the first plateau and why this plateau is smeared out
when $h$ approaches the critical point (figure~\ref{fig1}(b)). The second decay
of the magnetization is related to the lowest excitation. In the ferromagnetic phase,
the lowest energy gap vanishes exponentially with the system size as $\epsilon_0(L) \sim \exp(-L/\xi)$, $\xi$ 
being the correlation length, thus the corresponding time scale $t_d\sim 1/\epsilon_0$ grows exponentially with $L$,
as shown in the inset of figure~\ref{fig1}(a). At the critical point, the lowest excitation vanishes as 
$\epsilon_0(L) \sim \exp(-A L^{1/2})$ (see (\ref{psi})); thus, different relaxation behavior is expected for
a quench to the critical point and this is discussed in the following subsection.  

\subsubsection{Critical final state}


\begin{figure}[t]
\begin{center}
\includegraphics[width=6.3cm]{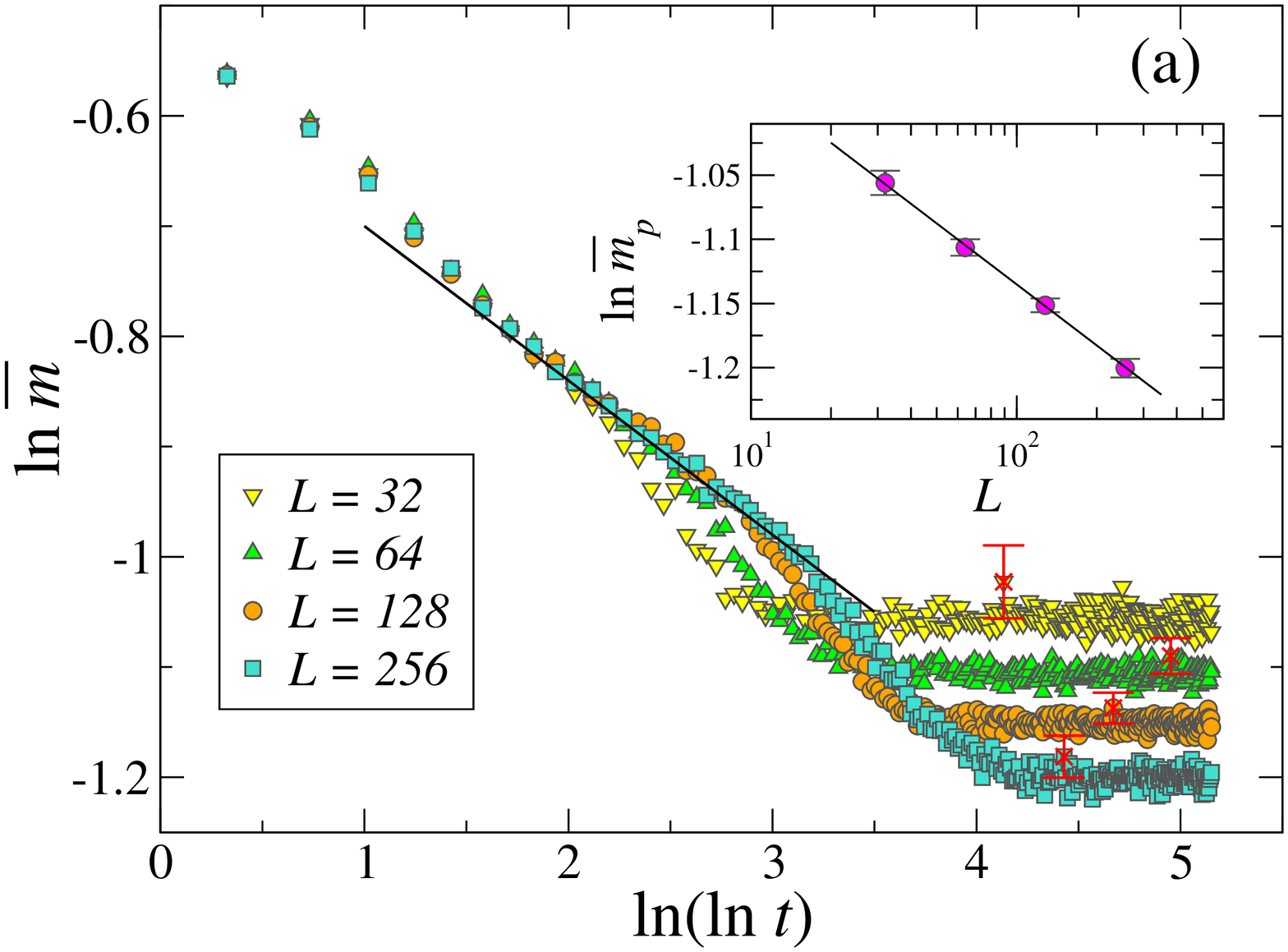}
\includegraphics[width=6.3cm]{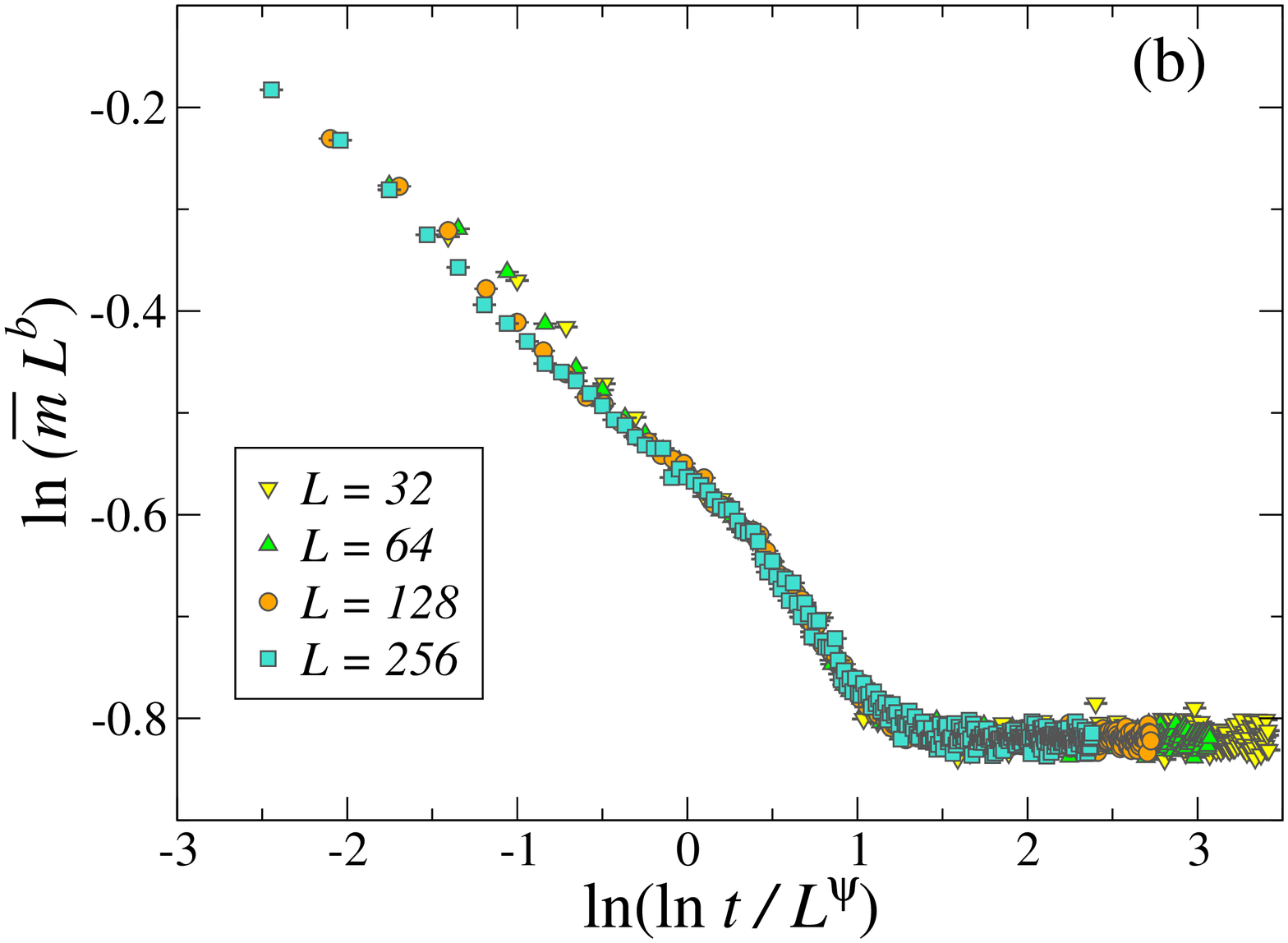}
\end{center}
\caption{
Relaxation of the magnetization after a quench from a fully ordered state to the critical point.
(a) The straight line indicating the decay in the transient region has a slope $a=0.14$.
The long-time limiting values as a function of $L$ are shown in the inset in a log-log plot;
the slope of the straight line is given by $b=0.069$. The largest errors in each data set are given by the error bars (red).
(b) A scaling plot of the data in (a),
with the parameters $b=0.069$ and $\psi=0.5$.}
\label{fig2}
\end{figure}


Relaxation of the average magnetization from the fully ordered initial state to
the critical point is shown in figure~\ref{fig2} for various chain lengths.
Here the first plateau observed in a ferromagnetic final state is missing, and
instead there is a transient region showing a linear relation between 
$\ln \overline{m}(t)$ and $\ln(\ln t)$ with a slope $a\approx 0.14$, corresponding to
a logarithmically show decay:
\be
\overline{m}(t) \sim (\ln t)^{-a}\;.
\label{a}
\ee
This transient regime terminates at a delay time, $t_d(L)$, which is an increasing function of $L$.
For $t>t_d(L)$ the magnetization decays faster and then saturates to a plateau with large-time 
limiting value $\overline{m_p}(L)$, which has a power-law $L$-dependence as
\be
\overline{m_p}(L) \sim L^{-b}\;;
\label{b}
\ee
the exponent $b$ is estimated as $b\approx 0.069$ in the inset of figure~\ref{fig2}(a).  Since the
two expressions in (\ref{a}) and (\ref{b}) should match at $t=t_d(L)$, we
arrive at the relation: $\ln t_d(L) \sim L^{b/a}$.  Comparing with
(\ref{psi_ne}), we have $\psi_{\mathrm{ne}}=b/a$, agreeing, within error bars,
with the RSRG-X prediction: $\psi_{\mathrm{ne}}=\psi=0.5$.  This result is
illustrated in figure~\ref{fig2}(b) in which the scaled magnetization, $L^b
\overline{m}(t)$ is plotted against the scaling combination $\ln t/L^{\psi}$;
with the exponents measured in figure~\ref{fig2}(a) we obtain an excellent scaling
collapse. Here we comment on the logarithmic decay described in (\ref{a}).  A similar
logarithmic decay is also present in the XX-chain with bond disorder, where a
larger exponent $a=2$ is found~\cite{Vosk_13,sirker}.  The critical states of
the random-bond XX-chain and the random transverse-field Ising chain are both
governed by an infinite-randomness fixed point in SDRG language.  The slower
decay (corresponding to a smaller exponent $a$) for the random transverse-field
Ising chain is due to effective spin clusters formed during renormalization. In
the semi-classical picture~\cite{Rieger_11}, relaxation of the order parameter is
contributed by quasi-particles created during the quench; a spin is
flipped if the trajectory of a quasi-particle crosses the given site at a
later time. For the relaxation of a larger cluster more quasi-particles are
needed, which takes more time.  The magnetic moment of an effective  spin
cluster of the random transverse-field Ising chain scales as $\mu \sim
L^{d_f}$, with a fractal dimension $d_f>0$.  The process of formation of spin
clusters is missing in the SDRG procedure for the XX-chain, where only random
singlets are formed, corresponding to $d_f=0$. The different SDRG procedures
for the Ising chain and the XX-chain lead to different functional forms of the
decay.


\begin{figure}[t]
\begin{center}
\includegraphics[width=6.3cm]{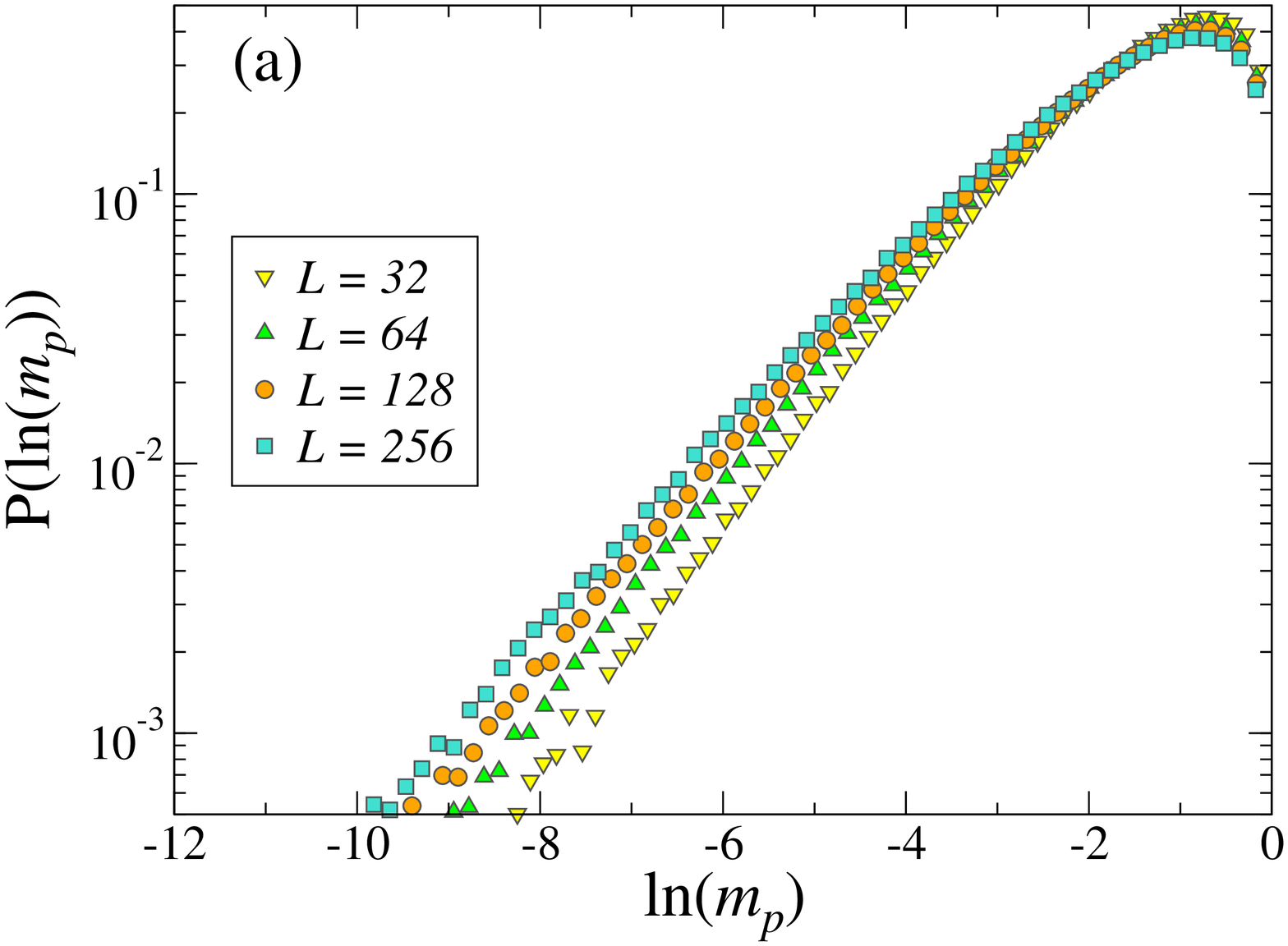}
\includegraphics[width=6.3cm]{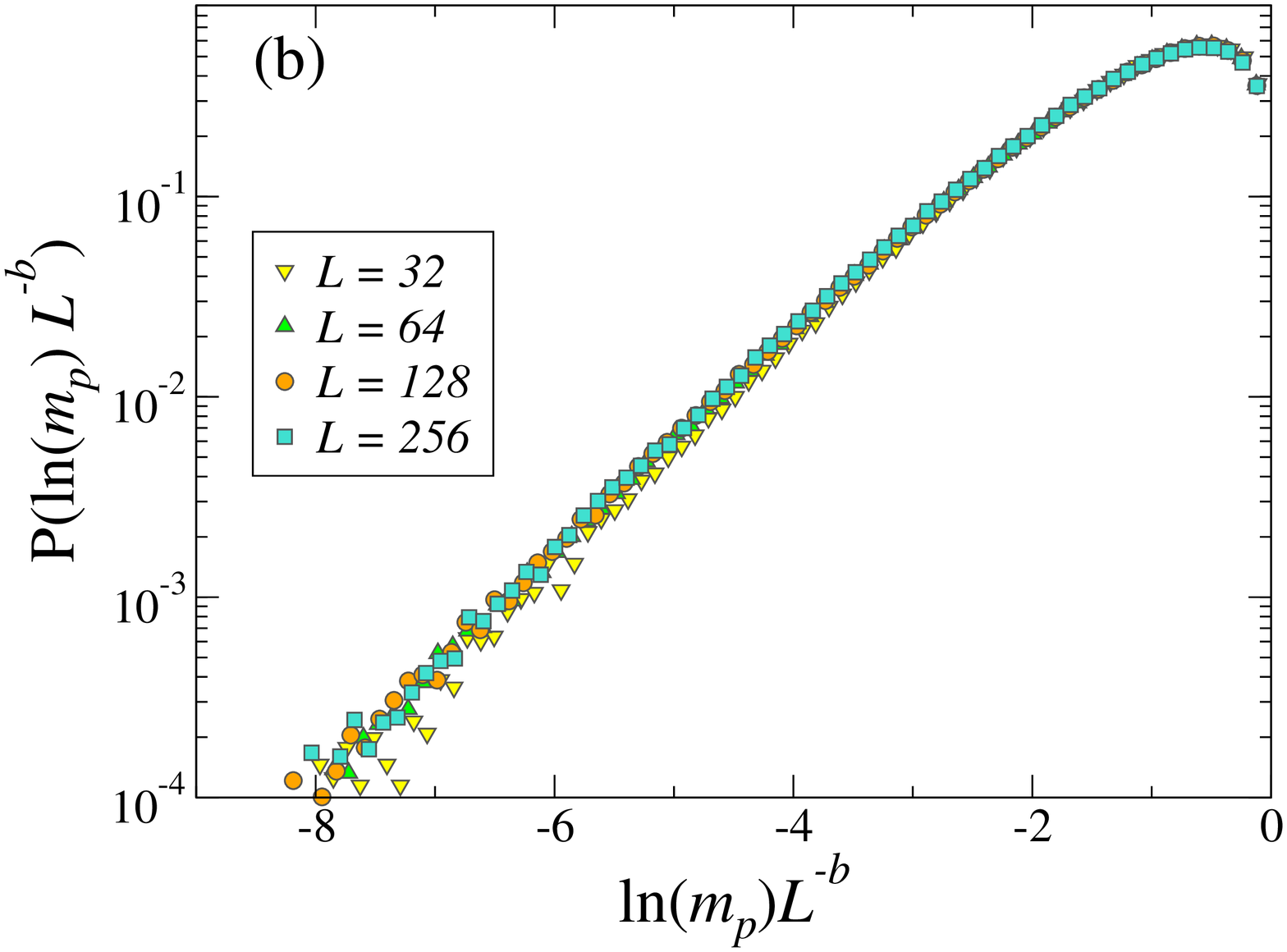}
\end{center}
\caption{
(a) Distribution of the large-time limiting values $m_p$ of the dynamical magnetization after a quench 
from a fully ordered state to the critical point for different chain lengths $L$. 
(b) A scaling plot for the data in (a) with $b=0.069$.
}
\label{fig3}
\end{figure}


In a random system the distributions of some observables are also of interest. 
Here we focus on the distributions of the large-time limiting values $m_p$ of the magnetization. 
In figure~\ref{fig3}(a) we plot the distribution of the logarithmic values $\ln(m_p)$ for
$t\gtrsim\exp(\exp(4.5))$ for different system sizes.
The distribution of the logarithmic variable is broad and becomes broader 
with increasing $L$. A good scaling collapse can be obtained in terms of the scaled variable $y=\ln(m_p) L^{-\alpha}$, 
thus 
\be
P_L(\ln m_p)=L^{-\alpha}\tilde{P}\bigl(\ln(m_p) L^{-\alpha}\bigr)\;,
\ee
as illustrated in figure~\ref{fig3}(b), with an estimated exponent
$\alpha=0.069$ equaling the exponent $b$ in (\ref{b}) for the scaling of the
average value. Furthermore, the typical value $m_p^{\mathrm{typ}}$ is
exponentially small and scales as $m_p^{\mathrm{typ}} \sim \exp(-CL^{\alpha})$,
where $C$ is a constant;
comparing with the average value in (\ref{b}), it is then obvious that the
average value is determined by rare events (or samples), in which the
large-time limiting value of the magnetization is $m_p(L)={\cal O}(1)$.
Consequently, the average value of $m_p(L)$ is determined by  
the behavior of the distribution function in the limit of $y \to 0^{-}$,
which is assumed to be in a power-law form:
\be
\tilde{P}(y) \sim (-y)^{\chi}\;,
\label{chi}
\ee
with $\chi=0$ in our case shown in figure~\ref{fig3}(b). 
The average of $m_p$ is then given by
\beqn
\overline{m_p}&=&L^{-\alpha}\int {\rm d} m_p\, \tilde{P}\bigl(\ln(m_p) L^{-\alpha}\bigr) \nonumber \\
&\sim& L^{-\alpha}\int {\rm d} m_p\, \bigl[\ln(m_p) L^{-\alpha}\bigr]^{\chi} \sim L^{-\alpha(1+\chi)}\;;
\label{m_p_av}
\eeqn
thus, with $\alpha=b$ and $\chi=0$ we recover the relation $\overline{m_p}\sim L^{-b}$, as given in (\ref{b}).

\subsection{Relaxation from a fully disordered initial state}
Now we turn to the case in which the system is quenched from a fully
disordered initial state ($h_0=\infty$) to the critical point ($h=1$).  


\begin{figure}[t]
\begin{center}
\includegraphics[width=6.3cm]{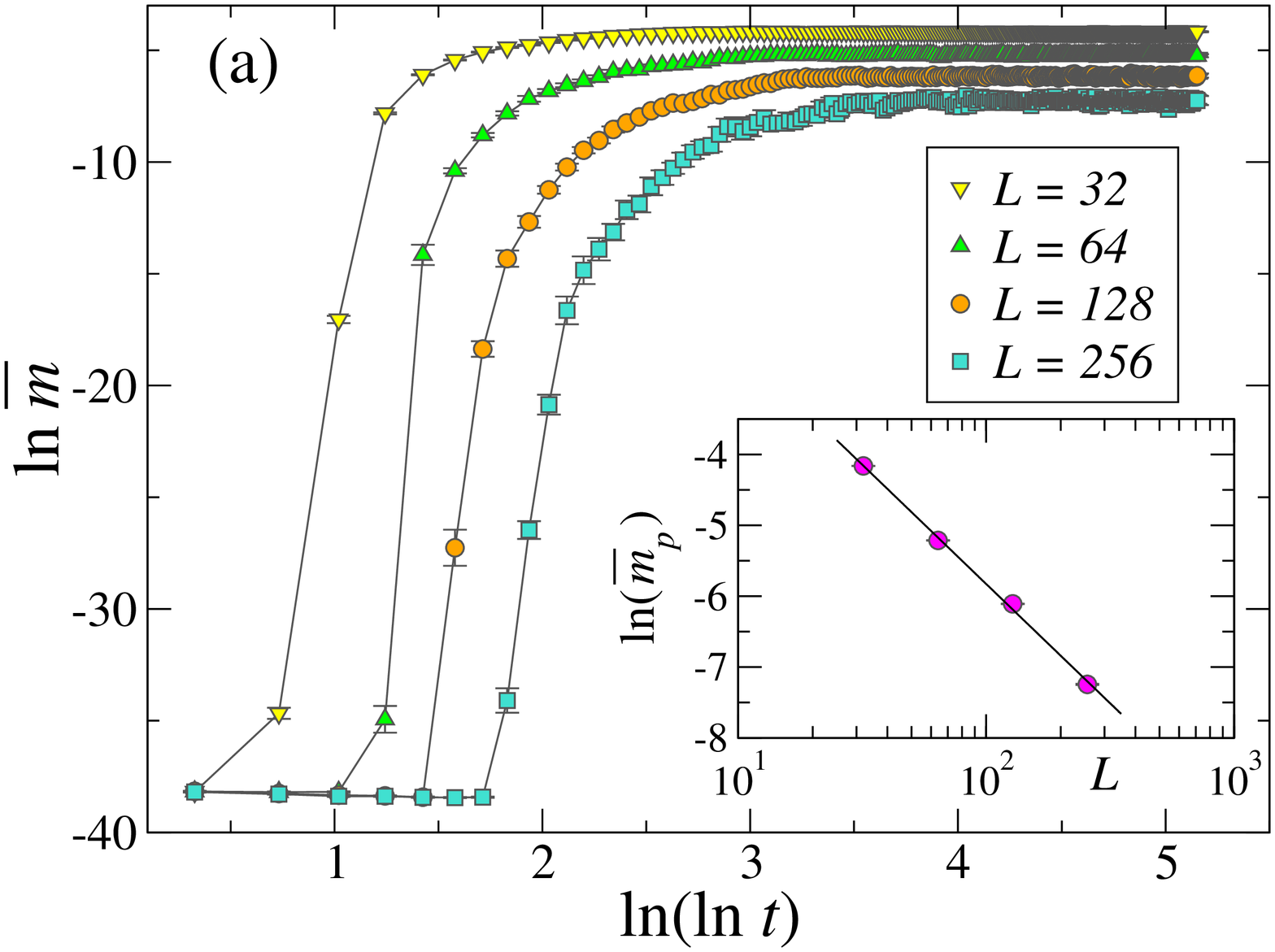}
\includegraphics[width=6.3cm]{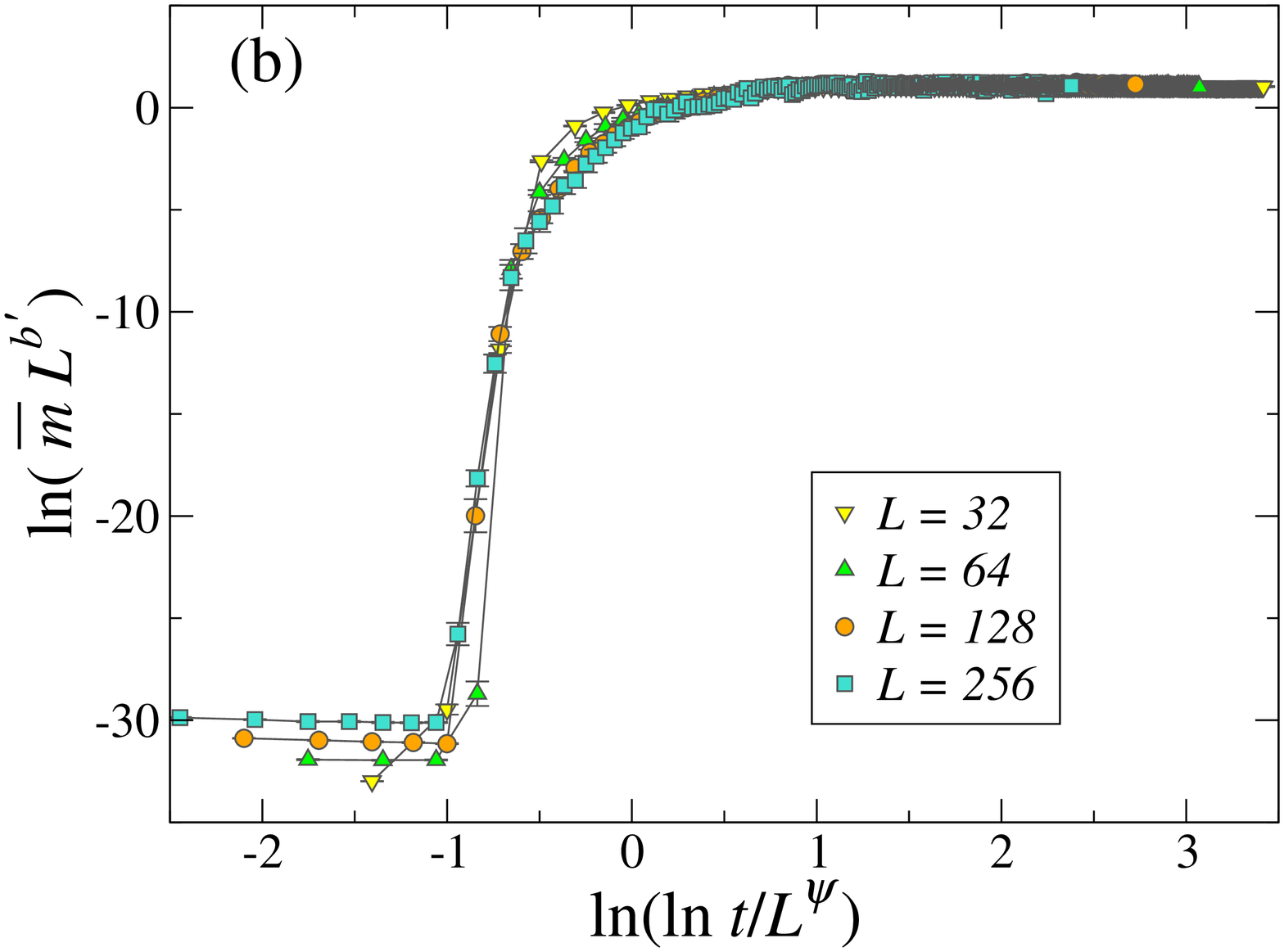}
\end{center}
\caption{
(a) Time-dependent average magnetization in finite chains after a
quench from a fully disordered state to the critical point. In the inset the
large-time limiting values are shown as a function of $L$ in a log-log plot. The
slope of the straight line is given by $b'=1.46$. (b) A scaling plot of the data in
(a) with the parameters $b'=1.46$ and $\psi=0.5$.
}
\label{fig4}
\end{figure}


Looking at the time-dependent average magnetization for various system sizes plotted in figure~\ref{fig4}(a), 
there is a period of time $0<t<t_d(L)$ just after the quench where the magnetization stays negligible and
practically corresponds to the initial magnetization before the quench. After this delay time $t_d$, which is
$L$-dependent, the average magnetization starts to increase rapidly and in the large-time limit it approaches a plateau,
the value of which has a power-law $L$-dependence:
\be
\overline{m_p}(L) \sim L^{-b'}\;,
\label{b'}
\ee
with $b'=1.46$, as measured in the inset of figure~\ref{fig4}(a).
Using the scaled variables $\overline{m} L^{b'}$ and $\ln t/L^{\psi}$, we achieve
a good data collapse for the time-dependent magnetization in the scaling plot shown in figure~\ref{fig4}(b).
The increase of the average magnetization in the disordered chain after the quench is unexpected since
semiclassical theory predicts a monotonous decrease of the dynamical magnetization in a homogeneous system.


\begin{figure}[t]
\begin{center}
\includegraphics[width=6.3cm]{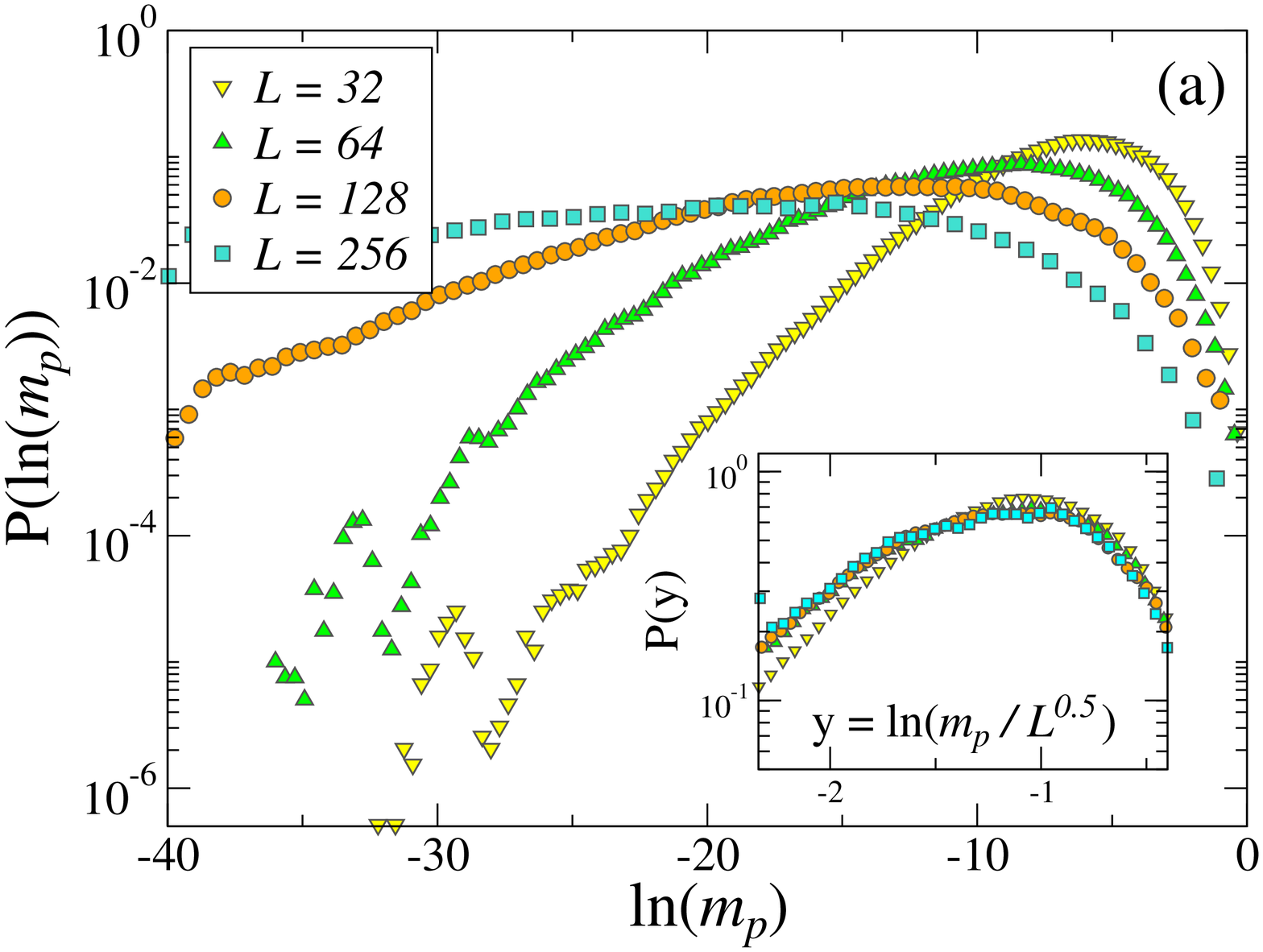}
\includegraphics[width=6.3cm]{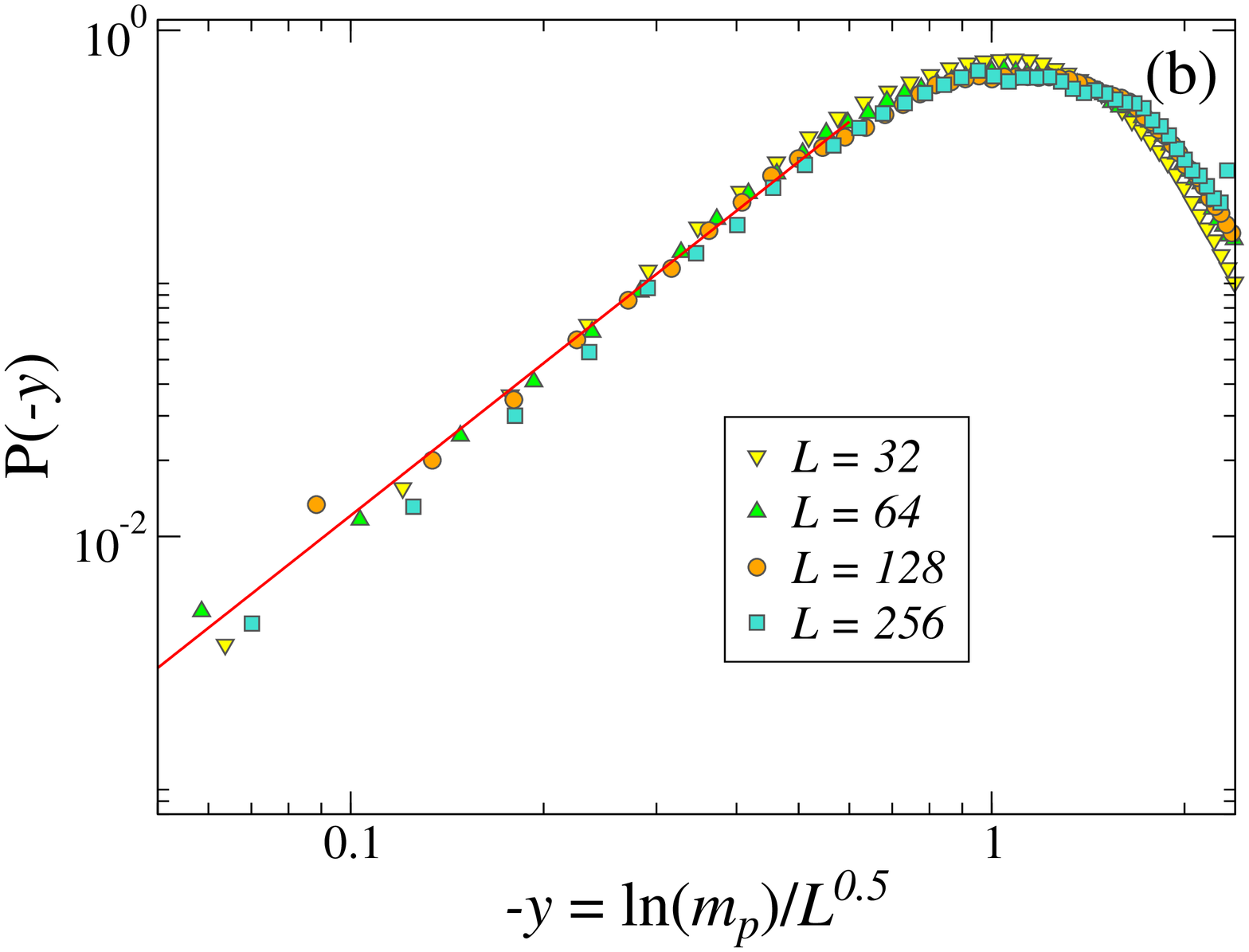}
\end{center}
\caption{
(a) Distribution of the large-time limiting values of the dynamical magnetization
after a quench from a fully disordered state to the critical point
calculated for different chain lengths; inset: a scaling plot, assuming $\ln m_p \sim L^{\alpha'}$ with $\alpha'=0.5$.
(b) The asymptotic behavior of the scaled distribution in terms of $y=\ln m_p/\sqrt{L}$ in the limit of $y\to 0^-$.
The red solid line corresponds to $P(y)\sim (-y)^2$.
}
\label{fig5}
\end{figure}


To understand the mechanism behind the increase of the average magnetization,
we have studied the distribution of the large-time limiting values
$m_p(L)$, which is presented in figure~\ref{fig5}(a).  The distribution is
extremely broad even on the logarithmic scale and it gets broader with
increasing system size; in terms of a scaling combination $y=\ln m_p
L^{-\alpha'}$ with $\alpha'=0.5$, the inset of figure~\ref{fig5}(a)  shows a
scaling plot with good data collapse. We can then conclude that the typical
value scales with $L$ as $m_p^{\mathrm{typ}} \sim \exp(-C\,L^{\alpha'})$ and is
much smaller than the average given in (\ref{b'}). Thus also in this quench
process the average value is determined by rare events, in which the large-time
limiting values of the magnetization are $m_p(L)={\cal O}(1)$. Similar to the
situation for a quench from an ordered state as discussed in
section~\ref{sec:ordered_c}, we have found  a power-law form $\tilde{P}(y) \sim
(-y)^{\chi'}$ for $y \to 0^{-}$ with an estimated exponent $\chi'=2.0$ (shown
in figure~\ref{fig5}(b)). Thus, we have found the size-dependence of the average
magnetization in (\ref{m_p_av}) as  $\overline{m_p}(L) \sim L^{-\alpha'(1+\chi')}=L^{-1.5}$, which agrees
with (\ref{b'}) within the error bars.

The scaling behavior of the large-time limiting value of the dynamical order
parameter can be better understood if the boundary site ($l=1$) of an open chain is
considered. The large-time limiting value of the {\it surface magnetization}, $m_p^s$, 
can be exactly calculated (see equation (16) in Ref.~\cite{Igloi_13}):
\be
m_p^s=\Phi_1(1)\sum_{j=1}^L \Phi_1(j) \Phi_1^{(0)}(j)\;.
\ee
Here:
\beqn
\Phi_1(j)&=&\Phi_1(1)\prod_{i=1}^{j-1}\frac{h_i}{J_i}\,, \nonumber \\
\Phi_1(1)&=&\left[1+\sum_{l=1}^{L-1}\prod_{j=1}^{l}\left(\frac{h_j}{J_j}\right)^2\right]^{-1/2}\,.
\eeqn
For $\Phi_1^{(0)}(j)$ in the large-$h_0$ limit we have
$\Phi_1^{(0)}(j)=\delta_{l_m,j}$, where $l_m$ is the position of the largest
transverse field in the sample~\cite{Phi_1^0}. Thus
\be
m_p^s=[\Phi_1(1)]^2 \prod_{j=1}^{l_m-1}\left(\frac{h_j}{J_j}\right)\;,
\ee
where $\Phi_1(1)=m^s_{\mathrm{eq}}$ is the equilibrium value of the surface
magnetization of the chain~\cite{peschel,Igloi_98}, evaluated in the final
state, i.e. with $h=h_c$. It is known that the typical value of $m^s_{\mathrm{eq}}$ 
at the critical point scales as $\exp(-CL^{1/2})$~\cite{Igloi_98,cecile} and the same
scaling form holds for $m_p^s$; thus the proper scaling combination for $m_p^s$ is
$y=\ln(m_p^s)L^{-1/2}$, which in the same form for bulk spins discussed above.
Concerning the scaling behavior of the average value $\overline{m_p^s}$, it is
dominated by rare events. As described in Ref.~\cite{Igloi_98}, 
there is a close analogy between the random transverse-field Ising chain and
a one-dimensional random walk: to a given sample
with a set of couplings $J_i$, and transverse fields $h_i$, $i=1,2,\dots L$,
one can assign a one-dimensional random walk which starts at the origin and
takes consecutive steps, the length of the $i$-th step being $\ln (h_i/J_i)$.
For a rare realization of $m^s_{\mathrm{eq}}$, the associated random walk has a surviving
character, i.e. it stays at positive position for all steps. Concerning $m_p^s$,
here for a rare realization both $m^s_{\mathrm{eq}}$ and the product
$\prod_{j=1}^{l_m-1}\left(\frac{h_j}{J_j}\right)$ should be of order of ${\cal O}(1)$; 
in the language of random walks, this means that the walk is surviving and returns after $l_m$ steps.
This particular event has the probability: ${\cal P}(l_m)\sim l_m^{-3/2}(L-l_m)^{-1/2}$ and its average
value scale as: $\sum_{l_m} {\cal P}(l_m)/L \sim L^{-3/2}$, which
means that $\overline{m_p^s}(L) \sim L^{-3/2}$. The numerically observed
scaling behavior for the bulk magnetization in (\ref{b'}) is similar to
this form.

\section{Conclusions}
\label{sec:disc}
In this paper we have studied numerically the time dependent magnetization of
the random transverse-field Ising chain after a quench. In order to obtain
accurate numerical results we have mainly performed matrix product operations with
multiple precision arithmetics in our calculations, instead of any
eigenvalue solver routine. In this way we obtained our finite-size
results up to length $L=256$ free from numerical instability. We note
that some problems with numerical instability using eigenvalue solvers
occurred when studying the entanglement entropy in large systems like $L=256$~\cite{Igloi_12,sirker}.

For a quench from a fully ordered initial state to a state within the
ferromagnetic phase, the average magnetization is shown to relax first to a
plateau value, followed by a second relaxation after a
delay time $t_{d} \sim \exp(cL)$ to a second plateau value. This second
relaxation is attributed to quasi-localized modes which are present in finite
systems in the ferromagnetic phase. Both plateau values are $h$-dependent, and 
finite because in the free-fermionic picture the emitted
quasi-particles are localized and travel only a finite distance, which 
reduces the initial order parameter by a finite fraction only.
This behavior is similar to that observed in the Aubry-Andr\'e model when the
quench is performed to the localized phase~\cite{Igloi_14}.

When the system is quenched from a fully ordered initial state to the critical
finale state, the relaxation of the average magnetization is logarithmically
slow (see (\ref{a})), and the large-time limiting value decays as a power of
$L$ (see (\ref{b})).  Between the time scale ($t$) and the length scale ($L$)
we have found the same relation as in equilibrium, $\ln t\sim\sqrt{L}$, 
as predicted by the RSRG-X approach~\cite{pekker,Vosk_13,Vosk_14}. We have
also studied the distribution of large-time limiting values of the
magnetization: the typical value decreases exponentially with the system size
as $\exp(-CL^{\alpha})$, more rapidly than the average value, which
decays as a power-law and is determined by rare realizations. 
The relaxation process after the quench can be explained by
the diffusion of quasi-particles in  a random
environment; the travel distance ($\ell$) of the quasi-particles
within time $t$ scales like $\ln(t)\sim \sqrt{\ell}$.
 
In a quench process from a fully disordered initial state to the critical
point, one observes a delay time, $t_d \sim \exp(CL^{1/2})$, in which the
average magnetization is negligible, and for $t>t_d$ a rapid increase of
the average magnetization toward  an asymptotic large-time limiting value,
which has a power-law $L$-dependence (see (\ref{b'})).  The distribution of the
large-time limiting magnetizations shows that the typical and the average
values scale differently, and the average is determined by rare events.
Qualitatively and even quantitatively similar behavior is found for the time
dependence of the surface magnetization, which has exact scaling arguments.
It has been explicitly shown that there are rare samples,
in which after a delay time a stationary surface magnetization of order of ${\cal O}(1)$ 
develops. These rare samples will then dominate the average
value. Such a phenomenon is similar to phase ordering dynamics in classical
systems~\cite{coarsening}, but not expected in closed homogeneous
quantum systems.

\ack{This work was supported by the Hungarian National Research Fund under Grants No.~K109577 and K115959,
and the Ministry of Science and Technology (MOST) of Taiwan under Grants No.~105-2112-M-004-002 and
104-2112-M-004-002. YCL also acknowledges support from the MOST under Grant No.~102-2112-M-002-003-MY3
and National Center for Theoretical Sciences (NCTS).} 

\appendix
\section{Time evolution of Majorana operators in quadratic fermionic systems}
\label{sec:app}
Let us consider a general Hamiltonian, $\cal H$, which is
quadratic in terms of fermion creation, $c_k^{\dag}$, and annihilation, $c_k$,
operators and is given for $t > 0$ as:
\be
{\cal H}=\sum_{k,l=1}^L\left[ c_k^{\dag}{  A}_{kl}c_l +\frac{1}{2} \left(
c_k^{\dag}{  B}_{kl}c_l^{\dag} + {\rm h.c.} \right)\right] \;.
\label{Hquad}
\ee
Here ${  A}_{kl} \equiv (\mathbf{A})_{kl}={  A}_{lk}$ and ${  B}_{kl}\equiv
(\mathbf{B})_{kl}=-{  B}_{lk}$ are real numbers, and $k,l$ are the sites of a
lattice. In the initial state, i.e. for $t<0$, the parameters of the initial Hamiltonian (${\cal H}^{(0)}$)
are different, say ${A}_{kl}^{(0)}$ and ${B}_{kl}^{(0)}$, and the ground state
of the initial Hamiltonian is denoted by $|\Psi_0^{(0)}\rangle$.

For $t>0$ the Heisenberg equation of motion for the operators $c_{k,H}(t)$ and $c_{k,H}^{\dag}(t)$ are
given by $\frac{\rm d}{{\rm d}t}c_{k,H}^{(\dag)}(t)=\imath[{\cal H},c_{k,H}^{(\dag)}(t)]$, i.e.
\beqn
\frac{\rm d}{{\rm d}t}c_{k,H}(t)=-\imath\sum_l[A_{kl}c_{l,H}(t)+B_{kl}c^{\dag}_{l,H}(t)]\,,\nonumber \\
\frac{\rm d}{{\rm d}t}c^{\dag}_{k,H}(t)=\imath\sum_l[A_{kl}c^{\dag}_{l,H}(t)+B_{kl}c_{l,H}(t)]\,,
\eeqn
which are linear since ${\cal H}$ is quadratic.
For the Majorana fermion operators $\check{a}_{2k-1}$ and
$\check{a}_{2k}$ at site $k$  as defined in (\ref{maj_def}) the time evolution is given by:
\beqn
\frac{\rm d}{{\rm d}t}\check{a}_{2k-1}(t)=-\sum_l[A_{kl}-B_{kl}](t)]\check{a}_{2l}(t)\;,\nonumber \\
\frac{\rm d}{{\rm d}t}\check{a}_{2k}(t)=\sum_l[A_{kl}+B_{kl}](t)]\check{a}_{2l-1}(t)\;,
\label{maj_deri}
\eeqn
which can be rewritten as
\be
\frac{\rm d}{{\rm d}t}\check{a}_{m}(t)=-\sum_{n=1}^{2L}M_{mn}\check{a}_{n}(t)\;.
\label{maj_deri1}
\ee
with a $2L \times 2L$ antisymmetric matrix ${\bf M}$ defined by
\beqn
M_{2k-1,2l}&=&-M_{2l,2k-1}=-(A_{kl}-B_{kl})\nonumber\;, \\
M_{2k,2l}&=&M_{2l-1,2k-1}=0\;.
\label{M}
\eeqn
The time dependent Majorana operators are related to the initial operators through
\be
\check{a}_{m}(t)=\sum_{n=1}^{2L}P_{mn}(t)\check{a}_{m}(0)\;,
\label{maj_rel}
\ee
with $P_{mn}(0)=\delta_{mn}$. Inserting (\ref{maj_rel}) into
(\ref{maj_deri1}) we obtain a set of differential equations for the time
evolution of the parameters:
\be
 \frac{{\rm d} P_{mn}(t)}{{\rm d} t}=\sum_{k=1}^{2L} M_{mk} P_{kn}(t);,
\ee
with the initial condition $P_{mn}(0)=\delta_{mn}$. 
The solution as a shorthand matrix notation is given in (\ref{P(t)}),
in which the exponential can be evaluated by using the spectral decomposition of ${\bf M}$.
Indeed, the eigenvalues of ${\bf M}$ are given by the energies of the free-fermionic
modes of ${\cal H}$ in (\ref{Hquad}), which can be seen by forming ${\bf
M}^2$ and comparing it with the results in Ref.~\cite{lieb61} . Details of the
calculation using the spectral decomposition of ${\bf M}$ are presented in the
appendix of Ref.~\cite{Igloi_13}.


\begin{thebibliography}{99}
\bibitem{Phi_1^0} It can be seen in Ref.\cite{Igloi_98}, by solving the eigenvalue problem in Eq.~(2.2) with $J_i=0$, $i=1,2,\dots,L$.
\end{thebibliography}
\end{document}